\documentclass[a4paper]{jpconf}
\usepackage{graphicx}
\begin{document}
\title{Domain walls with non-Abelian orientational moduli
}

\author{Minoru~Eto$^{a}$, 
Toshiaki~Fujimori$^b$, 
Muneto~Nitta$^c$,\\
Keisuke~Ohashi$^d$, 
and
 Norisuke~Sakai$^e$
}

\address{$^a$ Theoretical Physics Laboratory, RIKEN, Saitama, 351-0198, 
Japan, \\
$^b$ Department of Physics, Tokyo Institute of
Technology, Tokyo 152-8551, Japan, \\
$^c$ Department of Physics, and Research and Education center for 
Natural Sciences, Keio University, Hiyoshi 4-1-1, Yokohama,
Kanagawa 223-8521, Japan, \\
$^d$ Department of Physics, Kyoto University, Kyoto 
606-8502, Japan, \\
$^e$ Department of Mathematics, Tokyo Woman's Christian University, 
Tokyo 167-8585, Japan
}

\ead{$^a$ meto(at)riken.jp; 
$^b$ fujimori(at)th.phys.titech.ac.jp; \\
$^c$ nitta(at)phys-h.keio.ac.jp; 
$^d$ ohashi(at)gauge.scphys.kyoto-u.ac.jp; \\
$^e$ Speaker, sakai(at)lab.twcu.ac.jp
}

\begin{abstract}
Domain walls with non-Abelian orientational moduli are 
constructed in $U(N)$ gauge theories 
coupled to Higgs scalar fields with degenerate masses. 
The associated global symmetry is broken by the domain 
walls, resulting in the Nambu-Goldstone 
(and quasi-Nambu-Goldstone) bosons, 
which form the non-Abelian orientational moduli. 
As walls separate, the wave functions of the non-Abelian 
orientational moduli spread between domain walls. 
By taking the limit of Higgs mass differences to vanish, 
we clarify the convertion of wall position moduli 
into the non-Abelian orientational moduli. 
The moduli space metric 
and its K\"ahler potential of 
the effective field theory on 
the domain walls are constructed. 
We consider two models: a $U(1)$ gauge theory with several 
charged Higgs fields, and a $U(N)$ gauge theory with $2N$ 
Higgs fields in the fundamental representation.  
More details are found in our paper\cite{Eto:2008dm}. 
\end{abstract}

\section{BPS solitons}
Solitons are useful to build unified models with extra 
dimensions, and to provide all or part of nonperturbative 
effects. 
If a global symmetry of the theory is spontaneously broken 
by the presence of solitons, Nambu-Goldstone (NG) bosons 
come out and form (a part of) the moduli space of the soliton. 
In the case of non-Abelian global symmetry, the resulting 
massless modes can have non-Abelian orientational moduli. 
Quite often solitons have parameters, which are called 
moduli. 
If we promote these parameters to fields on the world 
volume of the soliton, they become massless fields in the 
low-energy effective field theory on the soliton.

Simplest of solitons is the domain wall which depends only 
on one spatial dimension, namely co-dimension one soliton. 
In order to have a domain wall, we need to have discrete 
vacua. 
As the simplest theory with two discrete degenerate vacua, 
let us consider a $1+1$ dimensional theory of real scalar 
field $\varphi$ with a double well potential 
\begin{equation}
{\cal L}=\partial_\mu \varphi \partial^\mu \varphi 
-\lambda (\varphi^2-v^2)^2, \qquad 
\lambda>0 . 
\label{eq:double_well_lag}
\end{equation}
If there is a field configuration connecting 
the two discrete degenerate vacua, 
$\varphi_+\equiv v$ and $\varphi_-\equiv -v$, we obtain 
a domain wall separating two vacua, as 
a kink, whose 
energy density is localized, resulting in the 
domain wall. 
The nontrivial boundary condition at infinity 
assures the topological stability of the configuration: 
topological charge is characterized by 
$\pi_0({\cal M})$. 
To obtain such a solution, let us assume a static 
configuration depending only on one spatial direction 
$y$, and form the following complete square of the 
energy density 
\begin{eqnarray}
{\cal E}
&=&(\partial_y \varphi)^2 +\lambda (\varphi^2-v^2)^2
\nonumber \\
&=& 
(\partial_y \varphi +\sqrt{\lambda}(\varphi^2-v^2))^2
+\partial_y\left[2\sqrt{\lambda}
\left(v^2 \varphi-{\varphi^3\over 3}\right)\right] .
\end{eqnarray}
This is called the Bogomol'nyi completion giving the 
lower bound of energy which is called the 
Bogomol'nyi bound 
\begin{equation}
\int_{-\infty}^{\infty} dy {\cal E} \ge 
\left[2\sqrt{\lambda}
\left(v^2 \varphi-{\varphi^3\over 3}\right)\right]_{-\infty}^{\infty} . 
\end{equation}

The Bogomol'nyi bound is saturated if and only if 
the following first order differential equation is satisfied: 
\begin{equation}
\partial_y \varphi +\sqrt{\lambda}(\varphi^2-v^2)=0, 
\end{equation}
which is called the Bogomol'nyi-Prasad-Sommerfield (BPS) 
equation for the domain wall. 
The BPS equation is easily solved to give the BPS 
domain wall solution 
\begin{equation}
\varphi=v \tanh (\sqrt{\lambda} v (y-y_0)) . 
\end{equation}
The BPS domain wall solution has a single modulus $y_0$, 
whose physical meaning is the position of the wall. 
This modulus can also be understood as a NG 
mode corresponding to the spontaneously broken 
translational symmetry. 
We can promote the real scalar to complex scalar, and 
add a fermion with appropriate interactions 
to make the system supersymmetric. 
Namely we can embed the theory (\ref{eq:double_well_lag}) 
into a supersymmetric theory with four supercharges. 
The BPS solution preserves the half of the supersymmetry 
in this embeded theory. 
This is a typical example of the BPS solitons which 
can be understood as the BPS state preserving 
a part of the supersymmetry in a supersymmetric theory. 

As another example of solitons, we can consider vortex 
in a theory with the Abelian gauge field $W_\mu$ coupled to 
a charged complex scalar field  $\phi$ 
\begin{equation}
{\cal L}=
-\frac{1}{4e^2} F_{\mu\nu}F^{\mu\nu}
 + {\cal D}^\mu \phi ({\cal D}_\mu \phi)^\dagger 
 -
\frac{\lambda}{4}\left(
\phi \phi^{\dagger}  - v^2\right)^2 , 
\end{equation}
where 
${\cal D}_\mu \phi =(\partial_\mu+i W_\mu)\phi$ denotes 
the covariant derivative, and 
$F_{\mu\nu}=\partial_\mu W_\nu - \partial_\nu W_\mu$ 
the field strength. 
The mapping from the infinity in the $x,y$ plane ($S^1$) to the 
vacuum manifold $|\phi|=v, S^1$ gives a 
topological charge $\pi_1(S^1)$ 
assuring the stability of the vortex configuration. 
\begin{equation}
k=-{1 \over 2\pi} \int d^2 x\ F_{12} 
\end{equation}
The position of a single vortex is again a modulus, since 
the vortex can be formed anywhere in the $x, y$ plane. 
If there are two or more vortices, the gauge field induces 
repulsive force between the vortices, and the scalar field 
induces attractive interactions\footnote{
This applies to vortices with vorticity of the same sign. 
If the signs of vorticities are opposite, both 
gauge field interactions and scalar interactions become 
attractive. 
}. 
When $e^2 = \lambda$, the two interactions cancel each other 
and there is no static force between vortices. 
Therefore vortices can be placed at anywhere relative to 
each other, and the relative positions become additional moduli. 
This critical case corresponds precisely to the case where the theory 
can be embedded into supersymmetric theory by adding fermions 
appropriately. 
The vortex solutions can then be understood as the BPS 
solitons preserving the half of the supersymmetry. 
If there are internal global symmetry which are broken by 
the presence of solitons, the NG modes emerge. 
In particular, if the global symmetry is non-Abelian, 
we can have non-Abelian orientational moduli for the soliton. 
These non-Abelian orientational moduli exhibits interesting 
properties similar to the D-branes in string theory.

\section{BPS equations for $U(N)$ gauge theories 
and the moduli matrix}\label{subsec:BPSeq}

We consider $U(N_{\rm C})$ gauge theory in 
space-time dimension $d=4+1$ with a real scalar field 
$\Sigma$ in the adjoint representation and 
$N_{\rm F}\,(>N_{\rm C})$ flavors of massive Higgs 
scalar fields in the fundamental representation, 
denoted as an 
$N_{\rm C} \times N_{\rm F}$ matrix $H$. 
Choosing the minimal kinetic term, we obtain 
\begin{eqnarray}
{\cal L} &=& {\cal L}_{\rm kin} - V, 
\label{eq:mdl:total_lagrangian}
\\ 
{\cal L}_{\rm kin} &=& 
{\rm Tr}\left(- {1\over 2g^2}F_{\mu\nu}F^{\mu\nu} 
+\frac{1}{g^2}{\cal D}_\mu \Sigma \, {\cal D}^\mu \Sigma 
+{\cal D}^\mu H \left({\cal D}_\mu H\right)^\dagger \right), 
\label{eq:mdl:lagrangian}
\end{eqnarray}
where the covariant derivatives and field strengths are 
defined as 
${\cal D}_\mu \Sigma=\partial_\mu\Sigma + i[W_\mu, \Sigma]$, 
\hspace{0.1cm}
${\cal D}_\mu H=(\partial_\mu + iW_\mu)H$, \hspace{0.1cm}
$F_{\mu\nu}=-i[{\cal D}_\mu,\,{\cal D}_\nu]$. 
Our convention for the space-time metric is 
$\eta_{\mu\nu} = {\rm diag}(+,-,\cdots,-)$. 
The scalar potential $V$ is given in terms of a 
diagonal mass matrix $M$ and a real parameter $c$ as 
\begin{eqnarray}
V&=& 
{\rm Tr}
\Big[
\frac{g^2}{4}
\left(c\mathbf{1}
-H  H^{\dagger} 
\right)^2 
+ (\Sigma H - H M) 
 (\Sigma H - H M)^\dagger 
\Big] . 
\label{eq:mdl:scalar_pot}
\end{eqnarray}

The 1/2 BPS equations for domain walls 
interpolating the discrete vacua can be obtained 
by usual Bogomol'nyi completion of the energy 
\begin{eqnarray}
E &=& \int_{-\infty}^{\infty} dy \, 
{\rm Tr} \left[ (\mathcal D_y H - HM + \Sigma H)^2 
+ \frac{1}{g^2} \left( {\cal D}_y \Sigma - \frac{g^2}{2} 
\left( c {\bf 1} - HH^\dagger\right) \right)^2 
+ c \, {\cal D}_y \Sigma \right] 
\\
&\geq& c \Big[{\rm Tr} \, \Sigma(\infty) 
- {\rm Tr} \, \Sigma(-\infty) \Big].
\end{eqnarray}
The first order differential equations for the 
configurations saturating this energy bound are of the 
form \cite{Isozumi:2004jc} 
\begin{equation}
\mathcal D_y H = HM - \Sigma H, \qquad \mathcal D_y \Sigma
= \frac{g^2}{2} \left(c{\bf 1} - HH^\dagger \right).
\label{eq:BPS}
\end{equation}
Here we consider static configurations 
depending only on the $y$-direction.

Let us solve these 1/2 BPS equations.
Firstly the first equation can be solved by \cite{Isozumi:2004jc}
\begin{equation
}
H=S^{-1}(y)H_0e^{My},\quad 
\Sigma+iW_y=S^{-1}(y)\partial_y S(y).
\label{eq:solBPS}
\end{equation
}
Here $H_0$, 
called the moduli matrix, 
is an $N_{\rm C} \times N_{\rm F}$ 
constant complex matrix of rank $N_{\rm C}$, 
and contains all the moduli parameters of solutions. 
The matrix valued quantity $S(y) \in GL(N_{\rm C},{\bf C})$ is determined by 
the second equation in (\ref{eq:BPS}) which can be converted 
to the following equation for $\Omega \equiv SS^\dagger$:
\begin{equation}
\frac{1}{cg^2} \bigl[ \partial_y(\Omega^{-1}\partial_y \Omega) \bigl] =
{\bf 1}_{N_{\rm C}}-\Omega^{-1}\Omega_0,\hspace{1cm}
\Omega_0 \equiv \frac{1}{c} H_0 e^{2My} H_0^\dagger .
\label{eq:master}
\end{equation}
This equation is called the master equation for domain walls. 
From the vacuum conditions 
at spatial infinities $y \rightarrow \pm \infty$, 
we can see that the solution $\Omega$ of the master equation should 
satisfy the boundary condition $\Omega \rightarrow \Omega_0$ 
as $y \rightarrow \pm \infty$. 
It determines $S$ for a given moduli matrix $H_0$ 
up to the gauge transformations $S^{-1} \rightarrow U S^{-1}, ~ U \in U(N_{\rm C})$ 
and then the physical fields can be obtained through (\ref{eq:solBPS}). 
Note that the master equation is symmetric under the following 
$V$-transformations
\begin{eqnarray}
H_0 \rightarrow VH_0 \quad 
{\rm and} \quad S(y) \rightarrow VS(y) \quad  {\rm with} \quad 
V \in GL(N_{\rm C}, {\bf C}), 
\label{eq:Vtransformation}
\end{eqnarray}
and if the moduli matrices are related by the $V$-transformations 
$H_0' = V H_0$, they give physically equivalent configurations. 
We call this equivalence relation as the 
$V$-equivalence relation and denote it as $H_0 \sim V H_0$. 
The solution of the master equation exists 
and unique for any given $H_0$ at least for 
the $U(1)$ gauge theory \cite{Sakai:2005kz}. 
 For $U(N)$ gauge theory the number of moduli parameters 
agrees with the the result of index theorem \cite{Sakai:2005sp}.

\section{Non-Abelian orientational moduli of walls in a $U(1)$ gauge theory
}
So far, domain walls with eight supercharges have been 
mostly considered in gauge theories with 
$U(1)$ gauge field \cite{Abraham:1992vb}--\cite{Sakai:2005kz}, 
$U(1)\times U(1)$ gauge field \cite{Eto:2005wf}, 
or $U(N)$ gauge fields \cite{Isozumi:2004jc}--\cite{Eto:2004vy} 
coupled to Higgs scalar fields with 
{\it non-degenerate} masses 
except for \cite{Shifman:2003uh,Eto:2005cc}. 
In the case of non-degenerate Higgs masses, 
the flavor symmetry is Abelian: $U(1)^{N_{\rm F}-1}$ 
and the symmetry of the vacua is also Abelian. 
As a result each domain wall carries a $U(1)$ 
orientational modulus \cite{Abraham:1992vb,Isozumi:2004jc};
The moduli space of a single domain wall is 
\begin{equation}
 {\cal W}^{k=1} \simeq {\bf R} \times S^1. \label{eq:mod-wall}
\end{equation} 
From this viewpoint,  
these domain walls should be 
called Abelian domain walls 
even when  the gauge symmetry of the Lagrangian 
is non-Abelian \cite{Isozumi:2004jc}--\cite{Eto:2005wf}. 

Let us see the non-Abelian orientational moduli in a simple example of 
the Abelian gauge theory coupled with the $N_{\rm F}=4$ 
Higgs fields.

\subsection{Vacua of the $U(1)$ gauge theory} 
\label{sc:simplest}

\begin{figure}[htb]
\begin{center}
\includegraphics[height=4.5cm]{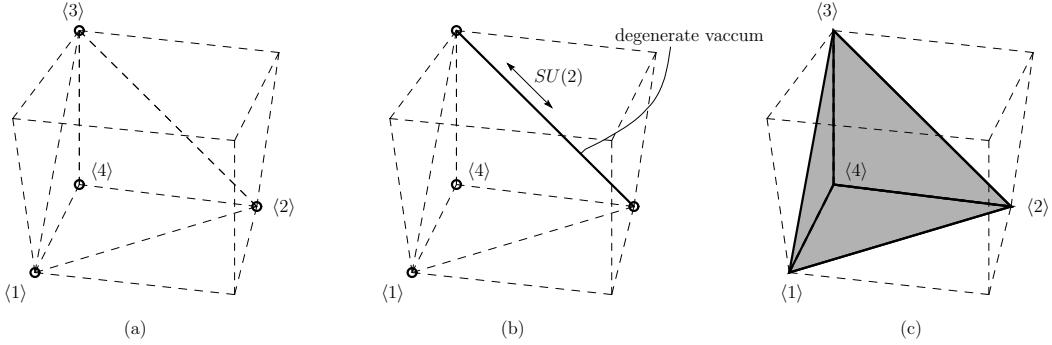}
\end{center}
\caption{\small
Vacua for various cases of mass configurations 
plotted in the three-dimensional space of Higgs 
fields $h_i^2, i=1,2,3$ with $\sum_{i=1}^{4}h_i^2=1$. 
(a) non-degenerate massive vacua 
(b) massive degenerate and non-degenerate vacua 
(c) massless vacuum.
}
\label{cp4}
\end{figure}

The massless vacuum manifold is $T^\star{\bf C}P^3$ 
where the base manifold is parametrized by
\begin{equation}
{\bf C}P^3 = \left\{HH^\dagger = c\right\}/U(1),\quad
H = \sqrt c \left(h_1,\ h_2,\ h_3,\ h_4 \right),
\end{equation}
where the quotient is the overall $U(1)$. 
The vacuum manifold is expressed as 
(the inside and the surface of)  
a triangular pyramid 
in the 3 dimensional space 
$(|h_1|^2, |h_2|^2, |h_3|^3)$, 
as shown in Fig.~\ref{cp4} (c). 
When the mass matrix 
containing a 
small 
parameter $\epsilon$ 
($0\le \epsilon \in {\bf R}$) 
\begin{eqnarray}
M = {\rm diag}
\left(
m,\ \frac{m\epsilon}{2},\  - \frac{m\epsilon}{2},\ -m
\right)
\label{eq:mass}
\end{eqnarray}
is turned on, the vacuum manifold is lifted except for 
four points and the flavor symmetry breaks from $SU(4)$ 
to $U(1)^3$.
These discrete vacua are the four vertices of the pyramid 
shown in Fig.~\ref{cp4} (a). 
We label those vacua as $\left<A\right>$ 
$(A=1,2,3,4)$. 
The vacuum expectation value (VEV) of the vacuum 
$\left<A\right>$ is $h_B = \delta_{AB}$.
Taking a limit of $\epsilon \to 0$, the second and the third 
Higgs fields become degenerate so that the flavor 
symmetry enhances 
from $U(1)^3$ to 
$U(1)^2 \times SU(2) \in SU(4)$. 
There are two isolated vacua 
and one degenerate vacuum 
${\bf C}P^1 \simeq SU(2)/U(1)$ 
represented by a line connecting $\left<2\right>$ and 
$\left<3\right>$ as shown by a thick line in 
Fig.\ref{cp4}(b).  
We denote this degenerate vacuum as $\left<\hbox{2-3}\right>$.

\subsection{Domain walls in the $U(1)$ gauge theory} 
\label{sc:simplest}

There exist domain wall solutions interpolating 
vacua in the model with fully or partially 
non-degenerate Higgs masses. 
In the case of $N_{\rm C}=1$, the moduli matrix and 
the $V$-equivalence (\ref{eq:Vtransformation}) 
take the form of 
\begin{equation}
H_0 = \left( \phi_1,\ \phi_2,\ \phi_3,\ \phi_4 \right)
\sim \lambda \left( \phi_1,\ \phi_2,\ \phi_3,\ \phi_4 \right),
\quad \lambda \in {\bf C}^*. 
\label{eq:N1Vequivalence}
\end{equation}
In terms of the moduli matrix the vacua $\left<A\right>$ 
are described by
$\phi_B = \delta_{BA}$ for $B=1,2,3,4$.
Since we want to consider the domain wall interpolating
the vacua $\left<1\right>$ and $\left<4\right>$ 
(passing by $\left<2\right>,\left<3\right>$ on the way),
the parameter $\phi_1$ and $\phi_4$
should not be zero while $\phi_2,\phi_3$ can become zero. 
So the moduli space corresponding
to the multiple domain walls which connect 
$\left<1\right>$ and $\left<4\right>$ is
\begin{equation}
 {\cal M} \simeq 
 \left( {\bf C}^2 \times ({\bf C}^*)^2\right) // {\bf C}^* 
 \simeq  {\bf C}^* \times {\bf C}^2, 
  \label{eq:moduli-simplest}
\end{equation}
where double slash denotes identification by 
the $V$-transformation.
Here the part ${\bf C}^* \simeq {\bf R} \times U(1)$ 
represents the translational modulus 
and the associated phase modulus.

\begin{figure}[htb]
\begin{center}
\includegraphics[width=95mm]{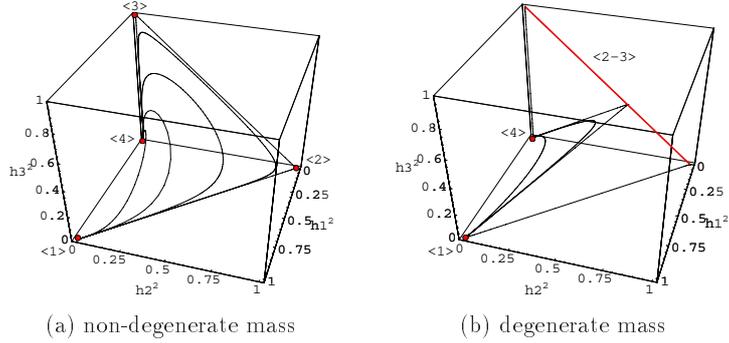}
\end{center}
\caption{\small
Domain wall trajectories in the target space ${\bf C}P^3$ 
for non-degenerate mass (a) and for degenerate mass (b). 
}
\label{cp3}
\end{figure}

When we take the gauge coupling $g$ to infinity, 
the model reduces to a nonlinear sigma model 
whose target space is the Higgs branch of 
the vacua $T^*{\bf C}P^3$ in the original theory. 
To make the discussion simple, we take this limit for a while.
One benefit to consider the nonlinear sigma model is 
that the BPS equations can be analytically solved. 
In fact the solutions are expressed as 
\cite{Isozumi:2004va} 
\begin{eqnarray}
H = \frac{1}{\sqrt \Omega_0} H_0e^{My}\quad
{\rm with}\quad
\Omega_0 \equiv H_0 e^{2My} H_0^\dagger. 
\end{eqnarray}
A domain wall solution corresponds to a trajectory 
connecting the vertex $\left<1\right>$ and 
$\left<4\right>$. Flows from $\left<1\right>$ to 
$\left<4\right>$ inside the pyramid
 are shown in Fig.~\ref{cp3}.

\begin{figure}[htb]
\begin{center}
\includegraphics[width=13.5cm]{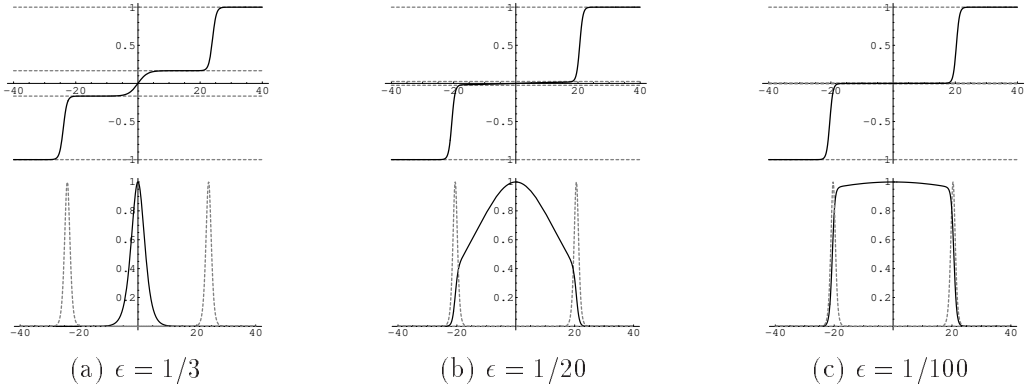}
\end{center}
\caption{\small
Configuration of $\Sigma$ 
(first row) 
and density of the K\"ahler metric 
of $\varphi_1$, $\varphi_2$
and $\varphi_3$ (second row). 
Moduli parameters are
$(\varphi_1,\varphi_2,\varphi_3)=(20,0,-20)$ and $m=1$.
}
\label{fig8}
\end{figure}

Physical meaning of the moduli parameters becomes much 
clearer by 
using the $V$-equivalence relation (\ref{eq:N1Vequivalence}) 
to fix the form of the moduli matrix as 
\begin{eqnarray}
H_0 =  \left( 1,\ e^{\varphi_1},\ 
e^{\varphi_1+\varphi_2},\ e^{\varphi_1+\varphi_2+\varphi_3} \right).
\label{eq:mm_3wall}
\end{eqnarray}
Furthermore, one may be visually able to see the ``kink" 
configuration in the profile of 
the field $\Sigma=(1/2)\partial_y \log \Omega_0$.
In the vacuum region $\left<A\right>$ the function $\Sigma(y)$ takes the value
$\Sigma = m_A$. Several solutions are shown in Fig.~\ref{fig8}.
The domain wall positions can be roughly read from the 
moduli matrix in Eq.~(\ref{eq:mm_3wall}) as 
\begin{eqnarray}
\frac{y_+}{L_+} = \varphi_1 + \varphi_1^*,\quad
\frac{y_0}{L_0} = \varphi_2 + \varphi_2^*,\quad
\frac{y_-}{L_-} = \varphi_3 + \varphi_3^*, \label{eq:3positions}
\end{eqnarray}
where $y_+$ is the position of the right wall and $y_0,y_-$ 
are the positions for the middle and the left walls, 
respectively. 
Here $L_{\pm,0}$ stands for the width of each wall 
\begin{eqnarray}
L_+ \equiv \frac{2}{m(2-\epsilon)}, \quad 
L_0 \equiv \frac{1}{m\epsilon},\quad 
L_- \equiv \frac{2}{m(2-\epsilon)}. 
\end{eqnarray}
This rough estimation is, of course, valid only for well 
separated walls whose positions are aligned as 
$y_- \ll y_0 \ll y_+$, see Fig.~\ref{fig8} (a). 
Each domain wall is accompanied by a complex moduli 
parameter $\varphi_i$ whose real part is related to the 
wall position and 
imaginary part 
is the $U(1)$ internal symmetry 
(the NG mode associated with the broken $U(1)$ 
flavor symmetry).  

To argue symmetry aspects of the moduli parameters, 
first let us consider a model which has completely 
non-degenerate masses and domain walls interpolating 
between those vacua. The 
global symmetry explicitly breaks
from $SU(4)$ to $U(1)^3 \subset SU(4)$. 
We take, as the unbroken 
global symmetries, $U_1(1),U_2(1)$ and $U_3(1)$
with generators 
${\rm diag}\left( 1, -1, -1, 1\right)$,
${\rm diag}\left( 1, 0, 0, -1\right)$,
and
${\rm diag}\left( 0, 1,-1, 0 \right)$
respectively.
Each vacuum $\left<A\right>$ preserves all of these symmetries.
However, once domain walls connecting those vacua appear, 
they break all or a part of these symmetries. For example, the moduli matrix $H_0 = (1,0,0,\phi_4)$ corresponding to a domain wall connecting two vacua 
$\left<1\right>$ and $\left<4\right>$ breaks $U_2(1)$ but still preserves 
$U_1(1)$ and $U_3(1)$. 
Here note that overall phase can be 
absorbed by the $V$-transformation (\ref{eq:N1Vequivalence}). 
Therefore the phase of the moduli parameter $\phi_4$ corresponds to nothing but the broken global symmetry $U_2(1)$.
This implies the NG mode localizes 
around the domain wall as we saw above.
For the moduli matrix $H_0=(1,\phi_2,0,\phi_4)$, 
which corresponds to two domain walls connecting three vacua 
$\left<1\right> \to \left<2\right> \to \left<4\right>$, 
the symmetry $U_3(1)$ in addition to $U_2(1)$ breaks while a combination of 
$U_1(1)$ and $U_3(1)$ is still preserved.
Moreover, when we turn on the third element in the moduli matrix as $H_0=(1,\phi_2,\phi_3,\phi_4)$,
the third vacuum region appears and then the configuration has three domain walls connecting four vacua 
$\left<1\right> \to \left<2\right> \to \left<3\right> \to \left<4\right>$.
In this case all of $U(1)^3$ are broken by the domain walls, so that 
corresponding three NG modes appear.
These three NG modes are described by
imaginary parts of $\log \phi$, 
which are combined with the three positions (\ref{eq:3positions}), 
to form three complex coordinates of the moduli space 
${\bf C}^2 \times {\bf C}^*$.

Next we consider a limit where the 
second and the third 
masses are degenerate 
($\epsilon\to0$ in the mass matrix (\ref{eq:mass})).
In the limit the global symmetry 
$U_1(1)\times U_2(1)\times U_3(1)$ is enhanced to 
$U_1(1)\times U_2(1)\times SU(2)$. 
At the same time, 
the degenerate vacuum $\left<\hbox{2-3}\right>$ appear 
instead of the two isolated vacua $\left<2\right>$ and
$\left<3\right>$ 
as shown in Fig.~\ref{cp4} (b). 
At the degenerate vacuum, $U_1(1),U_2(1)$ are preserved 
but $SU(2)$ is broken to $U_3(1)$. 
Therefore the degenerate vacuum $\left<2{\rm -}3\right>$ is 
$SU(2)/U_3(1) = {\bf C}P^1$. 
Non-vanishing $\phi_4\not=0$ causes 
the wall interpolating two vacua $\left<1\right> \to \left<4\right>$ 
and breaks only $U_2(1)$ again. 
Once the degenerate vacuum appears in the configuration such as 
two domain walls connecting vacua like 
$\left<1\right> \to \left<2{\rm -}3\right> \to \left<4\right>$,
the breaking pattern of the global symmetry becomes different 
from that in the case of fully non-degenerate masses. 
The moduli matrix $H_0 = (1,\phi_2,\phi_3,\phi_4)$ describes such domain walls. Note that the second and the third
elements break $SU(2)$ completely. The global symmetry $U_1(1) \times U_2(1) \times SU(2)$ are broken to $U(1)$ which 
is a mixture of $U_1(1)$ and $H \in SU(2)$. 
Emergence of the second wall and further $U(1)$-symmetry breaking 
are related to the facts that $|\phi_2|^2+|\phi_3|^2\not=0$ 
and $\phi_4\not=0$.
These facts imply that the modes corresponding to the 
two broken $U(1)$'s localize around the walls accompanied by the 
two position moduli and the mode corresponding to 
$SU(2)/H$ have support in a region around 
the degenerate vacuum $\left<2{\rm -}3\right>$.
We can count the number of the moduli parameters as follows.
Two real parameters $\{|\phi_2|^2+|\phi_3|^2,|\phi_4|^2\}$ 
correspond to the positions of the two walls 
whereas remaining four parameters 
correspond to the broken global symmetry 
$U_1(1) \times U_2(1) \times SU(2)/U(1)$. 
This is again consistent with
${\rm dim}_{{\bf R}} \left({\bf C}^2 \times {\bf C}^*\right)$.

In the Fig.~\ref{fig8} we showed domain wall configurations 
of the three domain walls connecting the four vacua. 
As the parameter $\epsilon$ 
decreases, 
the width of 
the middle domain wall connecting the vacua 
$\left<2\right>$ and $\left<3\right>$ becomes 
broad and the tension of the wall becomes small 
since they are proportional to $1/\epsilon$ 
and $\epsilon$, respectively. 
When the width of the middle wall becomes larger than the 
separation of two outside walls, 
$L_0 >
 y_+ - y_-$, 
we can no longer see the middle wall. 
The density of the K\"ahler metric for the moduli 
parameters $\varphi_1$, $\varphi_2$ and $\varphi_3$ in the strong 
gauge coupling limit are shown in the second row of 
Fig.~\ref{fig8}. 
The K\"ahler potential in the strong coupling limit 
is given by 
$K = c\int dy\ \log \Omega_0$ \cite{Eto:2006uw}. 
When three walls are well isolated as Fig.~\ref{fig8} (a), 
three modes corresponding to the moduli parameters 
$\varphi_1$, $\varphi_2$ and $\varphi_3$ 
are localized on the 
respective domain walls.  
As $\epsilon$ decreases, the density of the K\"ahler metric of 
$\varphi_2$ is no longer localized but is stretched 
between two outside domain walls. 
In the limit where $\epsilon\to 0$ the physical meaning of 
$\varphi_2$ as the position and the internal phase 
associated with the middle domain wall should be 
completely discarded. 
Instead, $\varphi_2$ gives the non-Abelian orientational moduli 
which comes 
from the 
${\bf C}P^1$ zero modes of the degenerate 
vacua $\left<2{\rm -}3\right>$. 
For fixed moduli parameters 
$\varphi_1, \varphi_2, \varphi_3$, 
the domain wall solution as a function of $y$ sweeps out 
a trajectory in the target space ${\bf C}P^3$. 
These domain wall trajectories are shown for various 
values of moduli parameters in Fig.~\ref{cp3}: 
the non-degenerate mass case (a) and the degenerate mass case (b). 
For the degenerate mass case, the trajectories 
do not go out from the triangular plane whose 
vertices are $\left<1\right>, \left<4\right>$ and one 
point on the edge between $\left<2\right>$ and 
$\left<3\right>$.

\section{The Generalized Shifman-Yung (GSY) Model}
\label{sc:GSY}

\subsection{$U(N)$ gauge theory 
}
\label{sc:strong-coupling}

Let us now consider non-Abelian gauge theory 
with degenerate masses of the Higgs fields. 
Previously considered model is the $U(2)$ gauge theory 
with four Higgs fields in the fundamental representation 
with the mass matrix $M = {\rm diag}(m,m,-m,-m)$ 
\cite{Shifman:2003uh,Eto:2005cc}, 
which we call the Shifman-Yung Model. 
The model enjoys a flavor symmetry 
$SU(2)_{\rm L}\times SU(2)_{\rm R}\times U(1)_{\rm A}$. 
It has been demonstrated that 
the coincident domain wall 
configurations break the flavor symmetry to $SU(2)_{\rm V}$ 
and the NG bosons corresponding to
$[SU(2)_{\rm L}\times SU(2)_{\rm R}\times U(1)_{\rm A}]/ SU(2)_{\rm V} 
\simeq U(2)$ 
appear in the effective action on the walls.

By generalizing the Shifman-Yung model, 
we consider the $U(N)$ gauge theory 
with $N_{\rm F}=2N$ Higgs fields in the fundamental representation 
whose mass matrix is given by  
\begin{eqnarray}
M ~=~ m {\sigma _3 \over 2} \otimes {\bf 1}_N 
~=~ {1\over 2}{\rm diag}(\overbrace{m,\cdots, m}^{N},
\overbrace{-m,\cdots,-m}^{N}).
\end{eqnarray}
This system has a non-Abelian flavor symmetry 
$SU(N)_{\rm L}\times SU(N)_{\rm R}\times U(1)_{\rm A}$. 
Since we have only two mass parameters $m$ and $-m$, possible 
vacua are classified by an integer $0 \le k \le N$: 
in the $k$-th vacua, there is a configuration in which 
$k$ flavors of the first half 
and $N-k$ flavors of the latter half take non-vanishing 
values and then $\Sigma $ and $H$ are 
\begin{eqnarray}
 \Sigma \big|_{\rm vacuum}&=&{1\over 2} \, {\rm diag}
(\overbrace{m,\cdots, m}^{k},\overbrace{-m,\cdots,-m}^{N-k}),
\nonumber  \\
H\big|_{\rm vacuum}&=&\sqrt{c}\left(
\begin{array}{cccc}
 {\bf 1}_k& {\bf 0}       & {\bf 0}_k &{\bf 0} \\
 {\bf 0}  & {\bf 0}_{N-k} & {\bf 0} &{\bf 1}_{N-k}
\end{array}
\right).
\label{eq:kthvacuum}
\end{eqnarray}
This vacuum is labeled as $(k,N-k)$. 
The flavor symmetry $SU(N)_{\rm L}$ is broken down to 
$SU(k)_{\rm C+L}\times SU(N-k)_{\rm L}\times U(1)_{\rm C+L}$,
and $SU(N)_{\rm R}$ 
is broken down to 
$SU(k)_{\rm R}\times SU(N-k)_{\rm C+R}\times U(1)_{\rm C+R}$, 
where $C+L, C+R$ denote the locking of the $L, R$ flavor 
symmetry with the $U(N)_{\rm C}$ color symmetry. 
Consequently 
the number of 
the discrete components of
the vacua is $N+1$ in this system. 
Therefore 
in this vacua there emerge 
$4k(N-k)$ NG modes, 
which parametrize 
the direct product of two Grassmann manifolds,
\begin{eqnarray}
G^{\rm L}_{N,k}\times G^{\rm R}_{N,k} . 
\label{eq:intra-vacua} 
\end{eqnarray}

Walls are obtained by interpolating between a vacuum 
at $y=-\infty$ and another vacuum at $y=\infty$. 
The boundary conditions at both infinities define topological 
sectors. 
For a given topological sector, we may find several walls. 
The maximal number of walls in this system is $N$, 
which are obtained for the following maximal topological 
sector 
\begin{eqnarray}
 H=\left\{\begin{array}{cc}
\sqrt{c}({\bf 1}_N,{\bf 0}_N)& {\rm at~} y=+\infty \\
\sqrt{c}({\bf 0}_N,{\bf 1}_N)& {\rm at~} y=-\infty \\
	    \end{array}\right. . 
\label{eq:bound_cod}
\end{eqnarray}
The unbroken symmetries of the vacua 
$(N,0)$ and $(0,N)$ which we consider 
as the boundary condition of domain walls here, are 
$SU(N)_{\rm C+L} \times U(1)_{\rm C+L} \times SU(N)_{\rm R}$ 
and  
$SU(N)_{\rm L} \times SU(N)_{\rm C+R} \times U(1)_{\rm C+R}$, 
respectively.  
In this case, the moduli matrix $H_0$
can be set into the following form without loss of generality:
\begin{eqnarray}
 H_0=\sqrt{c}({\bf 1}_N, e^\phi) \sim \sqrt{c}(e^{-\phi},\,{\bf 1}_N) ,
\end{eqnarray}
where $e^\phi $ is an element of $GL(N,{\bf C})$ and 
$\phi$ describes the moduli space of walls of this system, 
and the two forms are related by the $V$-transformation 
(\ref{eq:Vtransformation}). 
Therefore the moduli space of domain walls in the GSY model 
is 
\begin{equation}
 {\cal M} \simeq GL(N,{\bf C}) [= U(N)^{\bf C}]
\simeq \left[{\bf C}^* \times SL(N,{\bf C})\right]/Z_N .
 \label{eq:GSY-moduli}
\end{equation}
This moduli space admits the isometry 
\begin{equation}
 e^{\phi} \to e^{i \alpha} g_{\rm L} e^{\phi} g_{\rm R}^{\dagger} 
  \label{eq:isometry}
\end{equation}
with $(g_{\rm L},g_{\rm R}) \in SU(N)_{\rm L} \times SU(N)_{\rm R}$ 
and $e^{i \alpha} \in U(1)_{\rm A}$. 
This is because the domain wall solutions 
break the symmetry of the two vacua 
$(N,0)$ and $(0,N)$,  
$G = SU(N)_{\rm C+L} \times SU(N)_{\rm C+R} \times U(1)_{\rm C+L-R}$,  
down to its subgroup.

\subsection{Nambu-Goldstone (NG) modes and 
quasi-NG modes 
}\label{subsec:sym}

Note the fact that 
the global symmetry 
$G = SU(N)_{\rm L} \times SU(N)_{\rm R} \times U(1)_{\rm A}$ 
in (\ref{eq:isometry}) 
acts on the moduli space metric as an isometry 
whereas the complexified group 
$G^{\bf C} =SL(N,{\bf C})_{\rm L} 
\times SL(N,{\bf C})_{\rm R} \times {\bf C}^*$ 
acts on 
it transitively but not as an isometry.
Therefore $G^{\bf C}$ 
action may change the point in moduli space to 
another with a different symmetry structure. 
Since the symmetry of Lagrangian is $G$ but not $G^{\bf C}$ 
we can use only $G$ when we discuss the symmetry 
structure at each point in moduli space. 
General $\phi$ can be transformed by $G$ to 
\begin{equation}
 e^{\phi} = {\rm diag.} (v_1, v_2, \cdots, v_N) 
 \label{eq:general-vac}
\end{equation}
with $v_i$ real parameters.
When all $v_i$'s coincide (coincident walls), the unbroken 
symmetry is the maximal $H_{\rm max}=SU(N)_{\rm V}$, 
so we call it the symmetric point. 
There are $N^2$ massless NG bosons and 
$N^2$ quasi-NG bosons. 
When all $v_i$'s are different from each other (separated walls), 
$H_{\rm max} = SU(N)_{\rm V}$ is further broken down 
to 
$H_{\rm min} = U(1)_{\rm V}^{N-1}$.
Here 
the numbers of NG bosons and quasi-NG bosons
are $2N^2 - (N -1)$ and 
$N-1$, respectively. 
These $N-1$ quasi-NG bosons correspond to 
the $N-1$ parameters $v_i$ without the overall factor.
Therefore some of quasi-NG bosons 
at the symmetric point 
in the moduli space 
change to the NG bosons 
parametrizing
$H_{\rm max}/H_{\rm min} 
= SU(N)_{\rm V}/U(1)_{\rm V}^{N-1}$ 
reflecting this 
further symmetry breaking. 
When some $v_i$'s coincide, 
some non-Abelian groups 
are recovered: 
$H = U(1)_{\rm V}^r \times \prod U(n_i)_{\rm V}$.
Then the NG modes $H_{\rm max}/H$ are 
supplied from quasi-NG modes. 

The 
diagonal moduli parameters $v_i$ 
(quasi-NG bosons)
in Eq.~(\ref{eq:general-vac}) 
correspond to the positions of $N$ domain walls. 
When all domain walls are separated, 
the unbroken symmetry 
is $U(1)_{\rm V}^{N-1}$. 
When positions of $n$ domain walls coincide, $U(n)_{\rm V}$ 
symmetry is recovered. 
This phenomenon 
has a resemblance to 
the case of D-branes.  
However, there is a crucial difference: 
the symmetry in our case of domain walls is a global 
symmetry, whereas that of D-branes is a local 
gauge symmetry. 
However in the case of the $d=2+1$ wall world-volume, 
massless scalars can be dualized to gauge fields.
Shifman and Yung \cite{Shifman:2003uh} expected that 
the off-diagonal gauge bosons of $U(N)$ 
(which are originally the off-diagonal NG bosons of $U(N)$
before taking a duality) 
will become massive when domain walls are separated,   
in order to interpret domain walls as 
D-branes. 
However, our analysis shows that 
the off-diagonal NG bosons of $U(N)$
remain massless, and 
instead 
some of the quasi-NG bosons 
become NG bosons for further symmetry breaking 
with the total number of massless bosons unchanged  
as explained. 

We have obtained the low-energy effective Lagrangian explicitly 
for this model. 
We found the following. 
The wave functions of the 
NG boson for translation, for $U(1)^N$, and quasi-NG bosons 
are localized, and that other massless modes are extended 
between two domain walls, if walls are separated. 
Wave functions of all the massless modes become identical 
in the limit of coincident walls. 
These behaviors are different from D-branes, 
although there are some similarlities. 
More precise details of our results can be found in 
Ref.\cite{Eto:2008dm}. 

When we introduce complex masses for Higgs fields, 
domain wall junction or network emerge as $1/4$ BPS states 
\cite{Eto:2005cp}-\cite{Eto:2005mx}. 
In this case too, non-Abelain NG modes appear in the 
effective action, when some masses are degenerate 
\cite{Eto:2006bb}, \cite{Eto:2007uc}.

\ack
This work is supported in part by Grant-in-Aid for 
Scientific Research from the Ministry of Education, 
Culture, Sports, Science and Technology, Japan No.21540279,
No.21244036 (N.S.).

\appendix 
\setcounter{section}{1}
\section*{Appendix}

We will briefly give the derivation of our master equation 
(\ref{eq:master}). 
Let us first note an identity for variations $\delta_i, i=1,2$ 
of arbitary regular matrix $A$ 
\begin{eqnarray}
\delta_1(A^{-1} \delta_2 A)
&=& 
A^{-1}\delta_1 \delta_2 A 
-A^{-1} \delta_1 A A^{-1} \delta_2 A 
= 
A^{-1}(\delta_2 \delta_1 A A^{-1} 
-\delta_1 A A^{-1} \delta_2 A A^{-1})A
\nonumber \\
&=& 
A^{-1} \delta_2 (\delta_1 A A^{-1}) A
  \label{eq:identity}
\end{eqnarray}
For gauge invariant quantity $\Omega=S S^\dagger$, we obtain 
\begin{eqnarray}
&&\delta_1(\Omega \delta_2 \Omega^{-1})
= 
\delta_1(SS^\dagger \delta_2 (S^{-1\dagger}S^{-1})) 
=
\delta_1(SS^\dagger (\delta_2 S^{-1\dagger})S^{-1}
+S \delta_2 S^{-1}) 
\nonumber \\
&&
=S\left(
\delta_1(S^\dagger \delta_2 S^{-1\dagger}) 
+S^{-1}\delta_1S S^\dagger \delta_2 S^{-1\dagger} 
+S^\dagger \delta_2S^{-1\dagger} \delta_1 S^{-1}S 
+S^{-1}\delta_1(S \delta_2 S^{-1})S 
\right) S^{-1}
\nonumber \\
&&
=S\left(
\delta_1(S^\dagger \delta_2 S^{-1\dagger}) 
-\delta_1S^{-1}S S^\dagger \delta_2 S^{-1\dagger} 
+S^\dagger \delta_2S^{-1\dagger} \delta_1 S^{-1}S 
+\delta_2(\delta_1 S^{-1}S ) 
\right) S^{-1}
  \label{eq:variational_identity2}
\end{eqnarray}
By choosing both $\delta_1$ and $\delta_2$ to be dirivative in $y$, 
and noting that 
\begin{equation}
\Sigma={1\over 2}(S^{-1}\partial_yS +\partial_y S^\dagger S^{-1\dagger}), 
\quad
W_y={1\over 2i}(S^{-1}\partial_yS -\partial_y S^\dagger S^{-1\dagger}), 
\end{equation}
we obtain the master equation (\ref{eq:master}) for the domain walls. 
Choosing $\delta_1$ and $\delta_2$ as $\partial_z$ and 
$\partial_{\bar z}$ respectively, 
we can obtain the master equation for vortex.

\end{document}